\documentclass[preprint,5p]{elsarticle}
\usepackage{amssymb}
\usepackage{tikz}
\usepackage{braket}
\journal{Journal of Magnetism and Magnetic Materials}
\begin{document}
\begin{frontmatter}
\title{Spin-$1$ $J_1-J_2-J_3$ ferromagnetic Heisenberg model with an easy-plane crystal field on the cubic lattice: A bosonic approach}
\author{D. C. Carvalho$^1$, A. S. T. Pires$^{1}$, L. A. S. M\'ol$^{1}$}
\address{ $^1$Departamento de F\'\i sica, Instituto de Ci\^encias Exatas,
 Universidade Federal de Minas Gerais, C.P. 702, 
 30123-970 Belo Horizonte, MG - Brazil.
}

\begin{abstract}
We examine the phase diagram of the spin-$1$ $J_1-J_2-J_3$ ferromagnetic Heisenberg model with an easy-plane crystal field on the cubic lattice, in which
$J_1$ is the ferromagnetic exchange interaction between nearest neighbors,
$J_2$ is the antiferromagnetic exchange interaction between next-nearest neighbors and $J_3$ is the antiferromagnetic exchange interaction between 
next-next-nearest neighbors.  
Using the bond-operator formalism, we investigate the phase transitions between the disordered paramagnetic phase and the ordered ones. 
 We show that the nature of the quantum phase transitions change as the frustration parameters ($\frac{J_2}{J_1}$, $\frac{J_3}{J_1}$) 
are varied. The zero-temperature phase diagram exhibits second- and first-order transitions, depending on the energy gap behavior. Remarkably, we find
a disordered nonmagnetic phase, even in the absence of a crystal field, which is suggested to be a quantum spin liquid candidate.
We also depict the phase diagram at finite temperature for some values of crystal field and frustration parameters.
\end{abstract} 

\end{frontmatter}

\section{Introduction}
\label{intro}
The investigation of frustrated quantum spin models have attracted a lot of attention since Anderson proposed theoretically the 
existence of a non-magnetic ground state in a triangular lattice \cite{anderson}. Such a disordered state is known as a quantum spin liquid. 
From a theoretical point of view, the association between frustration and quantum fluctuations are believed to give rise to this novel quantum phase \cite{nature}.
While quantum fluctuations are introduced by the non-commutativity of quantum mechanics spin operators, frustration arises either from competition 
between different interactions, e.g. ferromagnetic and antiferromagnetic couplings \cite{cabra,pires}, or from geometry of the lattice, e.g. spins that interact 
via antiferromagnetic coupling on several lattices \cite{oitma, Wald}.

As a result, spin models that have both strong quantum fluctuations and frustration are candidates to exhibit quantum spin liquid phase. In fact, most of 
theoretical researches have been concentrated on spin-$1/2$ models in one- and two-dimensional lattices, since, in these cases, quantum fluctuations are enhanced
(see, for example, Refs.\cite{arXiv2015,richter,Majumda}). However, 
three-dimensional compounds have recently been suggested to present a quantum spin liquid phase \cite{PRL99,PRB90,tak}, 
which turns our attention to the theoretical investigation of three-dimensional models. 
 
In this paper, we study the phase diagram of spin-$1$ $J_1-J_2-J_3$ ferromagnetic Heisenberg model with an easy-plane crystal field on the cubic lattice, 
which is defined by the following Hamiltonian  
\begin{eqnarray}
\label{ham}
{\cal H}&=&-\frac{J_1}{2}\sum_{{\vec r},{\vec {{\delta}_{1}}}}{\vec S}_{\vec r}\cdot{{\vec S}_{{\vec r}+{\vec {\delta}_1}}}
+\frac{J_2}{2}\sum_{{\vec r},{\vec {{\delta}_{2}}}}{\vec S}_{\vec r}\cdot{{\vec S}_{{\vec r}+{\vec {\delta}_2}}}\nonumber\\
&+&\frac{J_3}{2}\sum_{{\vec r},{\vec {{\delta}_{3}}}}{\vec S}_{\vec r}\cdot{{\vec S}_{{\vec r}+{\vec {\delta}_3}}} +D\sum_{\vec r}\left(S_{\vec r}^{z}\right)^2,
\end{eqnarray}
where $\sum_{{\vec r},{\vec {{\delta}_{i}}}}$ sums over the first neighbors for $i=1$, over the second neighbors for $i=2$, and  
over the third neighbors for $i=3$. $J_1$ is the ferromagnetic exchange interaction between nearest neighbors,
$J_2$ is the antiferromagnetic exchange interaction between next-nearest neighbors and $J_3$ is the antiferromagnetic exchange interaction between 
next-next-nearest neighbors.
The last sum is over the total number of sites on the cubic lattice, $N$, and $D>0$ is the easy-plane crystal field that gives rise to a single-ion anisotropy.
${\vec S}_{\vec r}$ is the spin operator at site $\vec r$ with $S_{\vec r}^{z}$ taking the eigenvalues $-1,0,1$. 

To the best of our knowledge, the above Hamiltonian has not been examined yet. Only its counterpart model with all antiferromagnetic couplings ($J_1$, $J_2$,
$J_3$) has been treated in Ref. \cite{grif}, which is less interesting because in this case, $J_1$ and $J_3$ does not compete with each other. Thus the present
Hamiltonian (\ref{ham}) is much more suitable to seek quantum spin liquid candidates. In addition, compounds with both ferromagnetic and
antiferromagnetic exchange interactions have been reported in the literature (see Ref.\cite{cabra} and references therein), which indicates that the present
model with competitive interactions may be interesting not only from a theoretical perspective.

In order to study the phase diagram of this frustrated Heisenberg model, we use an analytical approach that has been successfully employed to describe transitions
from a gapped to a gapless phase, namely bond-operator formalism \cite{wang-wang,fquad,mucio,pires2}. In a few words, within the framework of the bond-operator
theory, the spin Hamiltonian is mapped into a Hamiltonian of non-interacting bosons, and phase transitions are located when the energy gap vanishes.

It should be pointed out that, in general, 
analytical approaches are more adequate to treat three-dimensional frustrated quantum spin systems than numerical methods, by reason of the limited amount of 
computational power. In particular, it is well known that quantum Monte Carlo suffers from the minus-sign problem 
which restricts its applicability: only non-frustrated quantum spin systems \cite{qmc}.

The present paper is organized as follows. In the next section we describe the bond-operator formalism and how phase transitions are characterized 
within this framework.
In Section \ref{res}, we show the results for the phase diagrams at finite temperature and at absolute zero as well. We also discuss the physical meaning of 
the obtained results. We close with some concluding remarks.

\section{Bond-operator Formalism}
\label{bof}
We employ the bond-operator formalism in order to investigate the phase diagram of the present model. This procedure was originally devised by Sachdev 
and Bhatt \cite{sach-bha}, for spin-$1/2$, and generalized by Wang {\it et al.} \cite {wang-shen,wang-wang}, for spin-$1$, some years later.
To begin we describe the method briefly below.

For spin-$1$, by introducing three boson operators, it is possible to represent the three eigenstates of $S^{z}$ as follows
\begin{eqnarray}
\label{nn-def}
 \Ket{1}&=&u^{\dagger}\Ket{v},\nonumber\\
\Ket{0}&=&t_{z}^{\dagger}\Ket{v},\nonumber\\
\Ket{-1}&=&d^{\dagger}\Ket{v}, 
\end{eqnarray}
where $\Ket{v}$ is the vacuum state from which bosons are created. These boson operators must also obey the local constraint  
\begin{equation}
  u^\dagger{u}+d^\dagger{d}+t_{z}^{\dagger}t_z=\hat1
\end{equation}
in order to keep the dimension of the local Hilbert space invariant.

Now, we can write the spin operators $S^x$, $S^y$ and $S^z$ in terms of these bosons operators
\begin{eqnarray}
\label{b-rep}
S^x&=&\frac{1}{\sqrt{2}}[(u^\dagger+d^\dagger)t_z+t_z^{\dagger}(u+d)],\nonumber\\
S^y&=&\frac{1}{\sqrt{2}i}[(u^\dagger-d^\dagger)t_z-t_z^{\dagger}(u-d)],\nonumber\\
S^z&=&u^\dagger{u}-d^\dagger{d}.
\end{eqnarray}
Substituting the above relations into Eq. (\ref{ham}), the Hamiltonian can be rewritten as a sum of four components
\begin{equation}
 {\cal H} = {\cal H}_1+{\cal H}_2+{\cal H}_3+{\cal H}_4,
\end{equation}
with
\begin{eqnarray}
 \label{h1}
{\cal H}_1&=&-\frac{J_1}{2}\sum_{{\vec r},{\vec {{\delta}_{1}}}}\bigg[{t}^2(d_{\vec r}^{\dagger}d_{{\vec r}+{\vec{\delta}_{1}}}
+u_{{\vec r}+{\vec{\delta}_{1}}}^{\dagger}
u_{\vec r} +u_{\vec r}d_{{\vec r}+{\vec{\delta}_{1}}}\nonumber\\
&+&d_{\vec r}^{\dagger}u_{{\vec r}+{\vec{\delta}_{1}}}^{\dagger}+H.c.)\nonumber\\
&+&\left(u_{\vec r}^{\dagger}u_{\vec r}-d_{\vec r}^{\dagger}d_{\vec r}\right)\left(u_{{\vec r}+{\vec{\delta}_{1}}}^{\dagger}u_{{\vec r}+{\vec{\delta}_{1}}}
-d_{{\vec r}+{\vec{\delta}_{1}}}^{\dagger}d_{{\vec r}+{\vec{\delta}_{1}}}\right)\bigg], \nonumber
\end{eqnarray}
\begin{eqnarray}
 \label{h2}
{\cal H}_2&=&\frac{J_2}{2}\sum_{{\vec r},{\vec {{\delta}_{2}}}}\bigg[{t}^2(d_{\vec r}^{\dagger}d_{{\vec r}+{\vec{\delta}_{2}}}
+u_{{\vec r}+{\vec{\delta}_{2}}}^{\dagger}
u_{\vec r} +u_{\vec r}d_{{\vec r}+{\vec{\delta}_{2}}}\nonumber\\
&+&d_{\vec r}^{\dagger}u_{{\vec r}+{\vec{\delta}_{2}}}^{\dagger}+H.c.)\nonumber\\
&+&\left(u_{\vec r}^{\dagger}u_{\vec r}-d_{\vec r}^{\dagger}d_{\vec r}\right)\left(u_{{\vec r}+{\vec{\delta}_{2}}}^{\dagger}u_{{\vec r}+{\vec{\delta}_{2}}}
-d_{{\vec r}+{\vec{\delta}_{2}}}^{\dagger}d_{{\vec r}+{\vec{\delta}_{2}}}\right)\bigg], \nonumber
\end{eqnarray}
\begin{eqnarray}
 \label{h3}
{\cal H}_3&=&\frac{J_3}{2}\sum_{{\vec r},{\vec {{\delta}_{3}}}}\bigg[{t}^2(d_{\vec r}^{\dagger}d_{{\vec r}+{\vec{\delta}_{3}}}
+u_{{\vec r}+{\vec{\delta}_{3}}}^{\dagger}
u_{\vec r} +u_{\vec r}d_{{\vec r}+{\vec{\delta}_{3}}}\nonumber\\
&+&d_{\vec r}^{\dagger}u_{{\vec r}+{\vec{\delta}_{3}}}^{\dagger}+H.c.)\nonumber\\
&+&\left(u_{\vec r}^{\dagger}u_{\vec r}-d_{\vec r}^{\dagger}d_{\vec r}\right)\left(u_{{\vec r}+{\vec{\delta}_{3}}}^{\dagger}u_{{\vec r}+{\vec{\delta}_{3}}}
-d_{{\vec r}+{\vec{\delta}_{3}}}^{\dagger}d_{{\vec r}+{\vec{\delta}_{3}}}\right)\bigg], \nonumber
\end{eqnarray}
\begin{eqnarray}
 \label{h4}
{\cal H}_4&=&D\sum_{\vec r}
\left(u_{\vec r}^{\dagger}u_{\vec r}+d_{\vec r}^{\dagger}d_{\vec r}\right)\nonumber\\
&-&\sum_{\vec r}\mu_{\vec r}\left(u_{\vec r}^{\dagger}u_{\vec r}+d_{\vec r}^{\dagger}d_{\vec r}+t^2-1\right),
\end{eqnarray}
where, by following the standard procedure \cite{wang-wang}, $t_z$ bosons have been condensed, i.e.
$\langle{t_z}\rangle=\langle{t_z}^{\dagger}\rangle=t$, and
a temperature-dependent chemical potential, $\mu_{\vec r}$, has been introduced to guarantee the single-site occupancy. $H.c.$ means 
Hermitian conjugate.
 
In order to diagonalize the Hamiltonian, it is necessary to decouple the four-operator terms of the Hamiltonian components ( 
${\cal H}_1$, ${\cal H}_2$ and ${\cal H}_3$) into product of two operators. To this end, we make a mean-field decoupling as done in Ref. \cite{wang-wang}.
For the component ${\cal H}_1$, for example, one obtains
\begin{eqnarray}
 \label{decoup}
\left(u_{\vec r}^{\dagger}u_{\vec r}-d_{\vec r}^{\dagger}d_{\vec r}\right)\left(u_{{\vec r}+{\vec{\delta}_{1}}}^{\dagger}u_{{\vec r}+{\vec{\delta}_{1}}}
-d_{{\vec r}+{\vec{\delta}_{1}}}^{\dagger}d_{{\vec r}+{\vec{\delta}_{1}}}\right)\approx\nonumber\\
\frac{1}{2}\left(1-t^{2}\right)\left(u_{\vec r}^{\dagger}u_{\vec r}+
u_{{\vec r}+{\vec{\delta}_{1}}}^{\dagger}u_{{\vec r}+{\vec{\delta}_{1}}}\right)\nonumber\\
+\frac{1}{2}\left(1-t^{2}\right)\left(d_{\vec r}^{\dagger}d_{\vec r}+
d_{{\vec r}+{\vec{\delta}_{1}}}^{\dagger}d_{{\vec r}+{\vec{\delta}_{1}}}\right)\nonumber\\
-p_1\left(u_{\vec r}d_{{\vec r}+\vec{\delta}_{1}}+ d_{\vec r}u_{{\vec r}+{\vec{\delta}_{1}}}+H.c.\right)\nonumber\\
-\frac{1}{2}(1-t^2)^2+2p_{1}^2,
\end{eqnarray}
where $p_1=\langle{d}_{\vec r}^{\dagger}u_{{\vec r}+\vec\delta_{1}}^{\dagger}\rangle=\langle{d}_{\vec r}u_{{\vec r}+\vec\delta_{1}}\rangle$. Similarly, one gets
$p_2=\langle{d}_{\vec r}^{\dagger}u_{{\vec r}+\vec\delta_{2}}^{\dagger}\rangle=\langle{d}_{\vec r}u_{{\vec r}+\vec\delta_{2}}\rangle$ and 
$p_3=\langle{d}_{\vec r}^{\dagger}u_{{\vec r}+\vec\delta_{3}}^{\dagger}\rangle=\langle{d}_{\vec r}u_{{\vec r}+\vec\delta_{3}}\rangle$, for components
 ${\cal H}_2$ and  ${\cal H}_3$, respectively. 
In addition, we make another approximation: the local constraint is replaced by a global constraint, that is, we assume that $\mu$ is site-independent. 

As a result, we arrive at a quadratic Hamiltonian involving boson operators. The next step in diagonalizing it is to take advantage of translational invariance by
using Fourier transformed operators
\begin{equation}
 d_{\vec r}=\frac{1}{\sqrt{N}}\sum_{\vec k}e^{-i{\vec k}\cdot{\vec r}}d_{\vec k},
\end{equation}
\begin{equation}
 d_{\vec r}^{\dagger}=\frac{1}{\sqrt{N}}\sum_{\vec k}e^{i{\vec k}\cdot{\vec r}}d_{\vec k}^{\dagger},
\end{equation}
\begin{equation}
 u_{\vec r}=\frac{1}{\sqrt{N}}\sum_{\vec k}e^{-i{\vec k}\cdot{\vec r}}u_{\vec k},
\end{equation}
\begin{equation}
 u_{\vec r}^{\dagger}=\frac{1}{\sqrt{N}}\sum_{\vec k}e^{i{\vec k}\cdot{\vec r}}u_{\vec k}^{\dagger}.
\end{equation}
However, this is not sufficient to diagonalize completely the Hamiltonian. Hence, we make use of a linear combination of Fourier transformed operators
\begin{eqnarray}
 \alpha_{\vec k}&=&\chi_{k}u_{\vec k}+\rho_{k}d_{-\vec k}^{\dagger},\nonumber\\
 \beta_{\vec k}&=&\chi_{k}d_{-\vec k}+\rho_{k}u_{\vec k}^{\dagger},
\end{eqnarray}
provided that $\chi_{k}^{2}-\rho_{k}^{2}=1$. This is known as Bogoliubov transformation.

Finally, we write the diagonal form of the Hamiltonian
\begin{equation}
\label{diag} 
{\cal H}=\sum_{\vec k}\omega_{k}\left(\alpha_{\vec k}^{\dagger}\alpha_{\vec k}+\beta_{\vec k}^{\dagger}\beta_{\vec k}\right)
+\sum_{\vec k}(\omega_{k}-\Lambda_{k}) + C,
\end{equation}
where 
\begin{equation}
\label{wk-heis}
 \omega_{k}=\sqrt{\Lambda_{k}^{2}-\Delta_{k}^{2}},
\end{equation}
\begin{eqnarray}
\label{vk}
\Lambda_k&=&\frac{1}{2}(1-t^2)(-z_1+\eta{z}_{2}+\alpha{z}_3) + g(k)t^2\nonumber\\
&+& D -\mu,
\end{eqnarray}
\begin{equation}
\label{dk}
\Delta_k = g(k)t^2 - F(k),
\end{equation}
\begin{equation}
 \label{gk}
 g(k)=-z_1\gamma_{k_1}+\eta{z}_{2}\gamma_{k_2}+\alpha{z}_3\gamma_{k_3},
\end{equation}
\begin{equation}
 \label{fk}
 F(k) = -z_1\gamma_{k_1}p_1+\eta{z}_{2}\gamma_{k_2}p_2+\alpha{z}_3\gamma_{k_3}p_3,
\end{equation}
\begin{equation}
 \gamma_{k_1} = \frac{1}{3}(\cos{k_x}+\cos{k_y}+\cos{k_z}),
\end{equation}
\begin{equation}
 \gamma_{k_2} = \frac{1}{3}(\cos{k_x}\cos{k_y}+\cos{k_x}\cos{k_z}+\cos{k_y}\cos{k_z}),
\end{equation}
\begin{equation}
 \gamma_{k_3} = \cos{k_x}\cos{k_y}\cos{k_z},
\end{equation}
\begin{eqnarray}
 C&=&\frac{N(1-t^2)^2}{4}\left(z_1-\eta{z_2}-\alpha{z_3}\right)+N\mu(1-t^2)\nonumber\\ 
 &+&N\left(-z_1{p_1}^2+\eta{z_2}{p_2}^2+\alpha{z_3}{p_3}^2\right).
\end{eqnarray}
Two frustration parameters have been defined, namely $\eta=\frac{J_2}{J_1}$ and $\alpha=\frac{J_3}{J_1}$, and we have fixed $J_1=1$.

We can derive the energy of the ground state from the Hamiltonian of non-interacting bosons (\ref{diag}) by setting 
$\alpha_{\vec k}^{\dagger}\alpha_{\vec k}=\beta_{\vec k}^{\dagger}\beta_{\vec k}=0$, since these operators count the number of bosons (excitations). Thus,
\begin{equation}
 E_{0}=C+\sum_{\vec k}(\omega_{k}-\Lambda_{k}).
\end{equation}
It is also straightforward to obtain the Gibbs free energy 
\begin{equation}
 G=E_{0}-\frac{2}{\beta}\sum_{\vec k}\ln[1+n(\omega_{k})],
\end{equation}
where 
\begin{equation}
n(\omega_{k})=\frac{1}{e^{\beta\omega_{k}}-1}
\end{equation}
and $\beta=\frac{1}{k_BT}$ is the Boltzmann factor.

By minimizing the Gibbs free energy, we obtain a set of coupled equations from which the phase diagram of the model can be examined,
\begin{equation}
\label{t}
 2-t^2=\frac{1}{N}\sum_{\vec k}\frac{\Lambda_k}{\omega_k}\coth\left(\frac{\beta\omega_k}{2}\right),
\end{equation}
\begin{equation}
\label{mi}
 \mu=\frac{1}{N}\sum_{\vec k}\frac{\Lambda_k-\Delta_k}{\omega_k}g(k)\coth\left(\frac{\beta\omega_k}{2}\right),
\end{equation}
\begin{equation}
\label{pi}
 p_i=-\frac{1}{2N}\sum_{\vec k}\frac{\Delta_{k}}{\omega_k}\gamma_{k_i}\coth\left(\frac{\beta\omega_k}{2}\right),
\end{equation}
with $i=1,2,3$. 

Taking the continuum limit, the summation over ${\vec k}$ can be replaced by a triple integral, then these integrals
are solved numerically by using the Gauss-Legendre method. We also apply the Newton-Raphson technique for solving these coupled equations.
\subsection{Phase Transitions}
\label{qpt}
The classical version of the Hamiltonian (\ref{ham}) with $D=0$ has two ordered phases at absolute zero:
\begin{enumerate}[(i)]
 \item A ferromagnetic phase (F) characterized by ${\vec k}_{F}=(0,0,0)$;
 \item  A collinear antiferromagnetic phase (CAF) characterized by ${\vec k}_{CAF}=(0,0,\pi)$ or ${\vec k}_{CAF}=(0,\pi,0)$ or ${\vec k}_{CAF}=(\pi,0,0)$.
\end{enumerate}
It should be mentioned that in contrast to the classical $J_1-J_2-J_3$ antiferromagnetic Heisenberg model, the collinear antiferromagnetic phase with 
${\vec k}=(0,\pi,\pi)$ is never stable in its ferromagnetic counterpart.

In order to verify the presence of these ordered phases in the quantum Hamiltonian (\ref{ham}), we must analyze the energy gap for both phases F and CAF.
According to the bond-operator formalism, phase transitions are obtained from the gapped phase (disordered) to the gapless phase (ordered) when the  
gap closes at the critical point. 

For the F order, the boson modes become gapless at ${\vec k}_{F}$ and therefore $\omega_{{\vec k}_{F}}=0$, which characterizes a phase transition between the 
disordered paramagnetic phase and the ferromagnetic phase. For the CAF order, the energy gap goes to zero at ${\vec k}_{CAF}$, then 
$\omega_{{\vec k}_{CAF}}=0$, which characterizes a phase transition between the disordered paramagnetic phase and the collinear antiferromagnetic phase 
%
\section{Results and Discussion}
\label{res}
In this section, we show numerical results obtained by solving the coupled equations (\ref{t}-\ref{pi}) in the continuum limit. 
Phase diagrams at zero temperature as well as at finite temperature are analyzed as a function of the parameters of the Hamiltonian (\ref{ham}).
\subsection{Quantum Phase Diagram ($T=0$)}
\begin{figure}[h!]
\begin{center}
\includegraphics[angle=0,width=8.5cm]{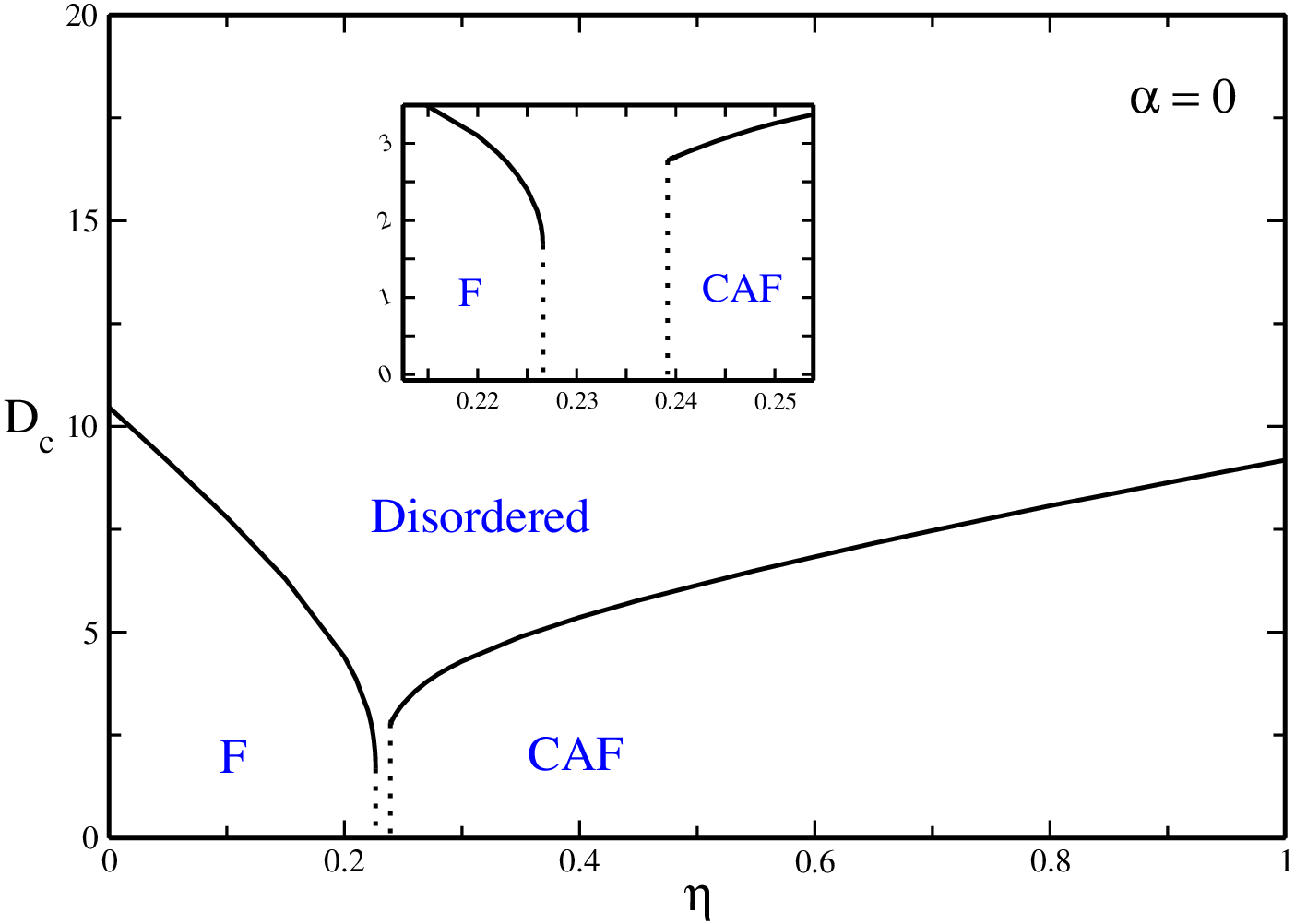} 
\caption{\label{alph0} (color online)
Critical crystal field, $D_c$, as a function of the frustration parameter, $\eta$, at zero temperature,
for $\alpha=0$. The continuous lines refer to second-order quantum phase transitions, 
and the dotted lines refer to first-order ones. The inset shows the low crystal field region on a finer scale.
}
\end{center}
\end{figure}

Figure \ref{alph0} shows the quantum phase diagram for $\alpha=0$. One notes the presence of two ordered phases below the critical lines, namely F and CAF. 
For $0\leq\eta<0.227$, the ferromagnetic phase is stable, while for $0.239<\eta\leq1$, the collinear antiferromagnetic phase is stable. A remarkable finding
is a narrow nonmagnetic phase between F and CAF phases along $\eta$-axis, for $0.227\leq\eta\leq0.239$, which is clearly depicted in the inset.
Such a gapped phase, which is absent in the corresponding classical model, is a quantum spin liquid candidate. Thus, our results indicate that even a 
three-dimensional model is able to host a quantum spin liquid phase, despite the fact that quantum fluctuations decreases with the increase of 
lattice dimensionality. Recently, a quantum spin liquid phase candidate
has also been suggested on the cubic lattice for the spin-$1/2$ $J_1-J_2-J_3$ antiferromagnetic Heisenberg model by using 
the variational cluster approach \cite{vojta}, which corroborates our finding.
%
\vspace{1cm}
\begin{figure}
\begin{center}
\includegraphics[angle=0,width=8.5cm]{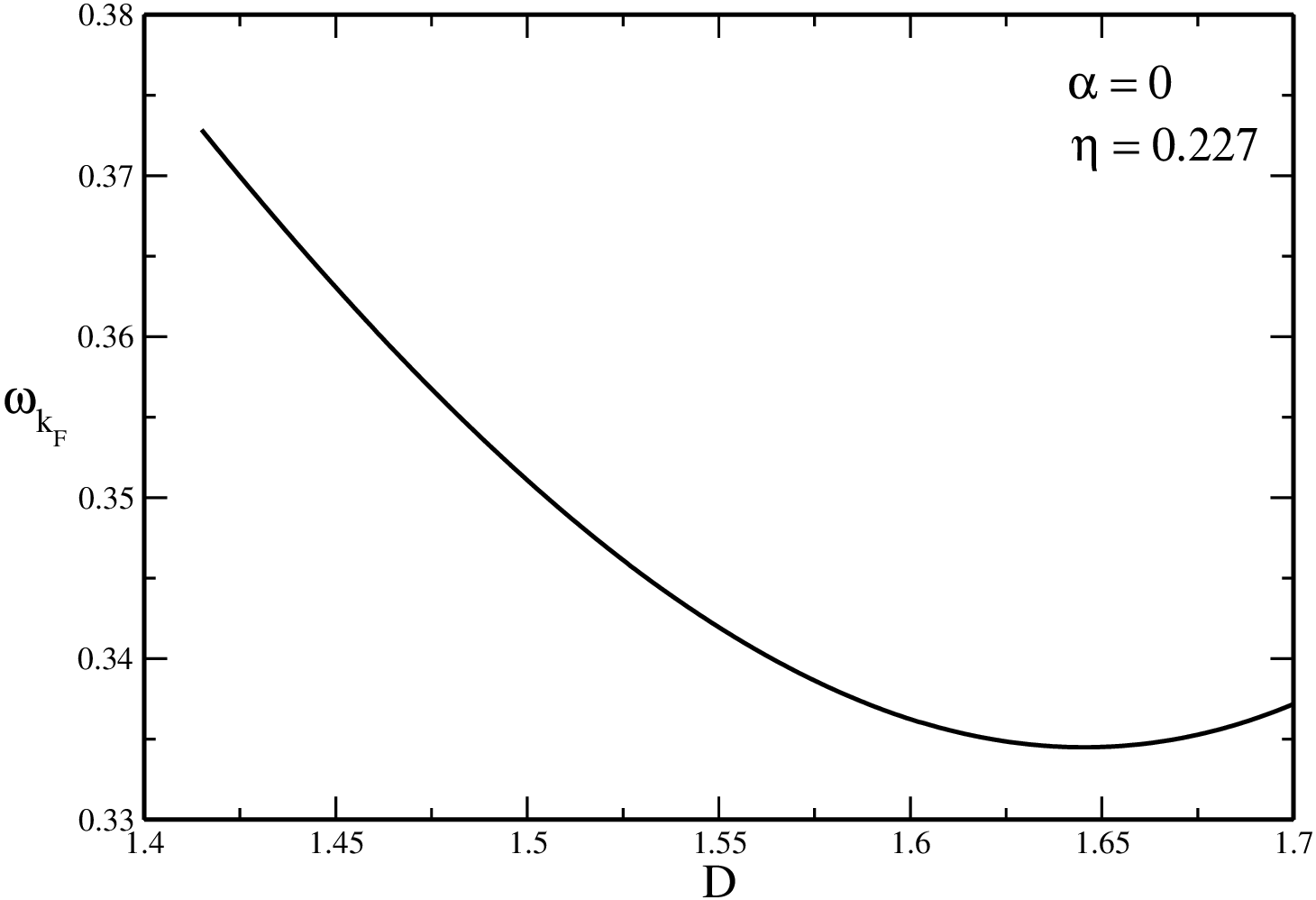} 
\caption{\label{gap} 
The energy gap of the ferromagnetic phase, $\omega_{{\vec k}_{F}}$, as a function of the crystal field, $D$, at zero temperature, for $\alpha=0$ 
and $\eta=0.227$.}
\end{center}
\end{figure}
%

One can also see that the nature of the phase transitions varies with $\eta$. For $\eta=0.227$ and $\eta=0.239$, the energy gap does not vanish continuously, 
as one would expect at a critical point, by contrast, it passes through a finite minimal value. This is shown in
Figure \ref{gap} for $\eta=0.227$. From $D=1.7$, we 
see clearly that the energy gap, $\omega_{{\vec k}_{F}}$, decreases with decreasing $D$, passes through a minimum, and then increases.  
Hence, we believe that the model undergoes first-order transitions for those values of $\eta$.

It is also worth analyzing the effect of the crystal field on the stability of the ordered phases. Note that the critical crystal field, $D_c$, for the F phase,
decreases with increasing $\eta$, while for the CAF phase $D_c$ increases. This can be understood as 
a result of the frustration that destroys the ferromagnetic order and favors the collinear antiferromagnetic one.

\begin{figure}[h!]
\begin{center}
\includegraphics[angle=0,width=8.5cm]{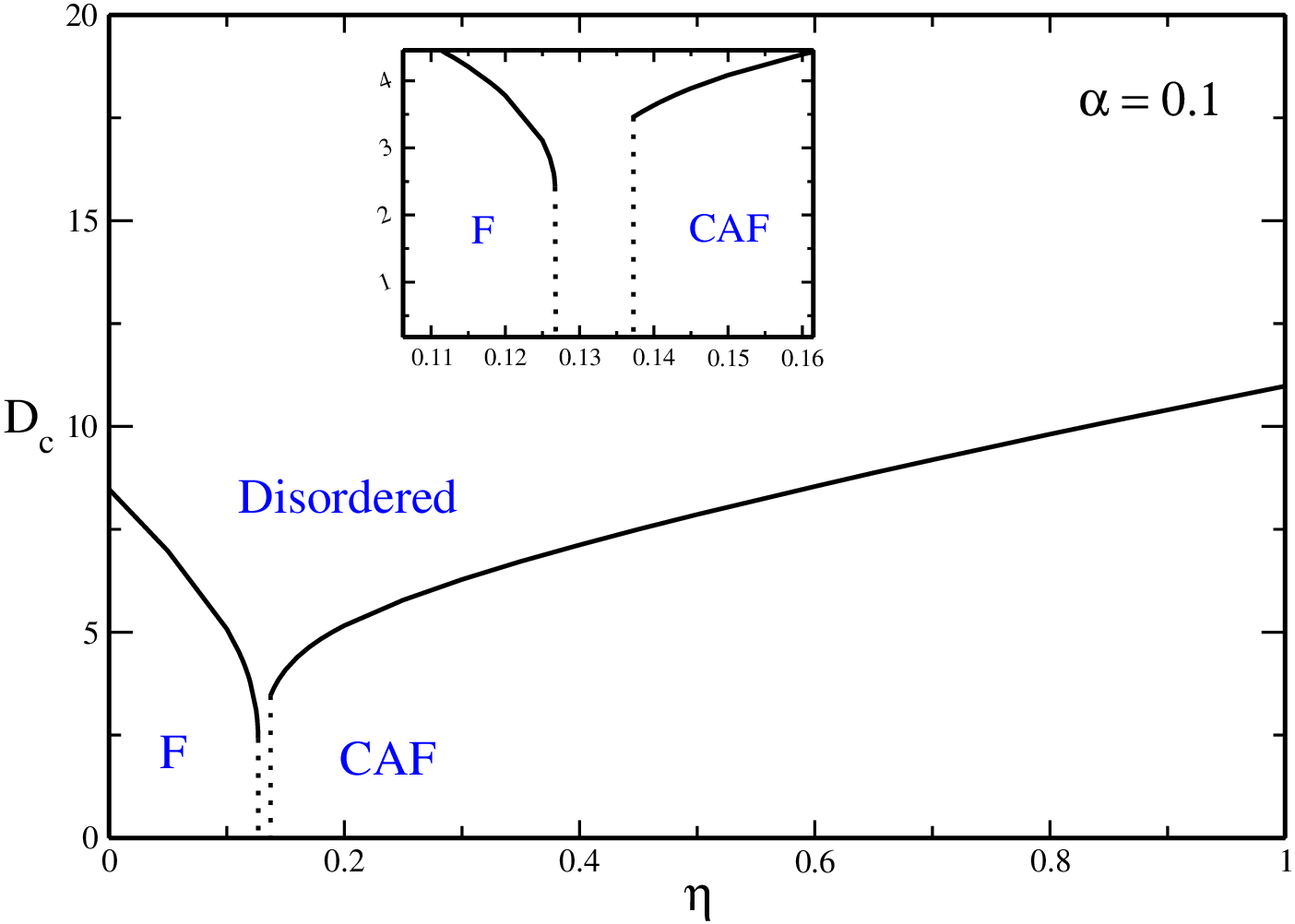} 
\caption{\label{alph01} (color online) 
The same as Fig. \ref{alph0} for $\alpha=0.1$.}
\end{center}
\end{figure}

A similar phase diagram is obtained by letting $\alpha=0.1$. As shown in Figure \ref{alph01}, when we include the third-neighbor 
coupling, $J_3$, the ordered phases, F and CAF, are still present as well as the magnetically disordered one. However, the $J_3$ interaction reduces the 
F phase and increases the CAF order. Considering $D<D_c$, the F phase is stable for $0\leq\eta<0.127$, while CAF phase is stable for $0.137<\eta\leq1$.

\begin{figure}
\vspace{0.05cm}
\begin{center}
\includegraphics[angle=0,width=8.5cm]{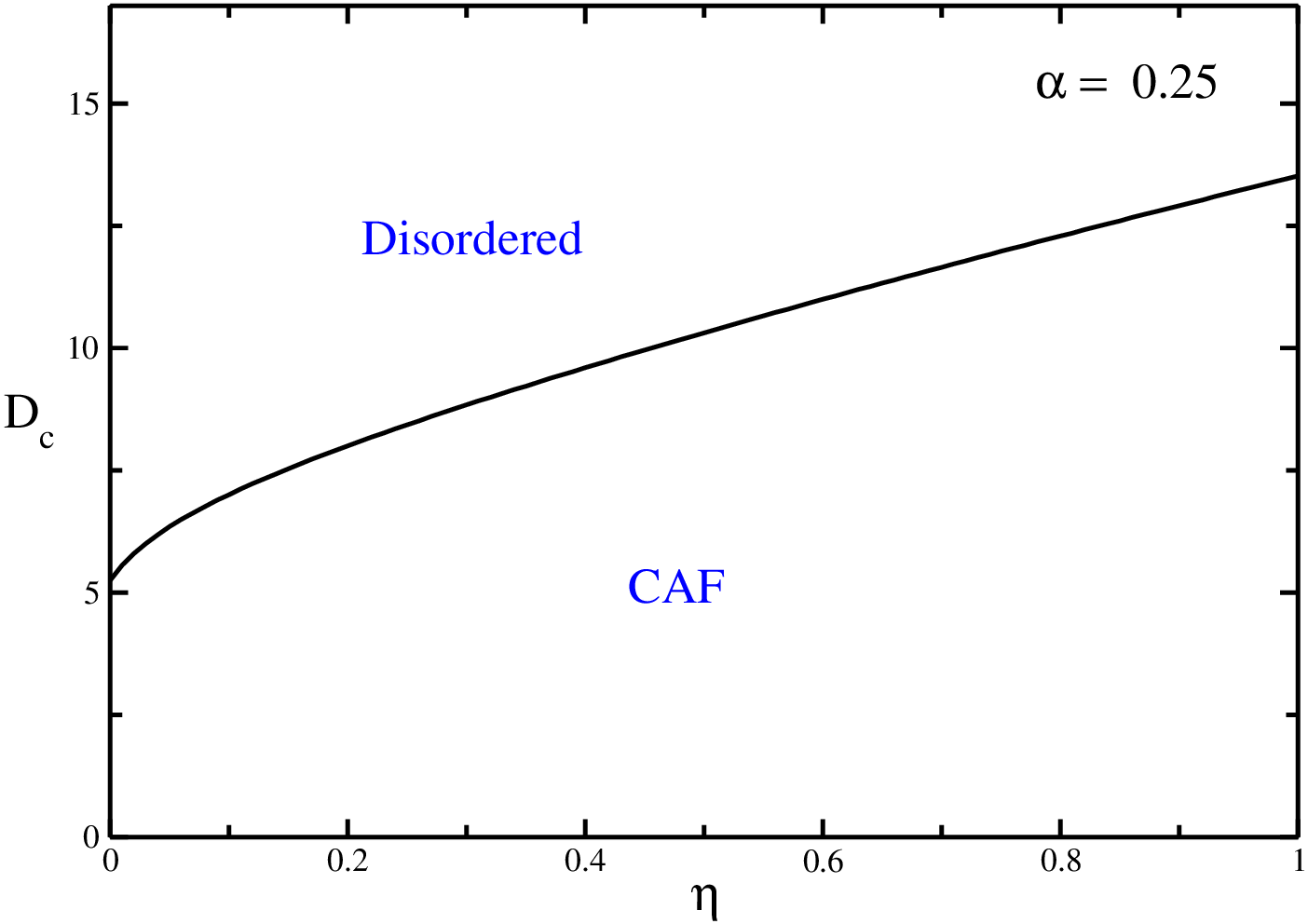} 
\caption{\label{alph025} (color online) 
Critical crystal field, $D_c$, as a function of the frustration parameter, $\eta$, at zero temperature,
for $\alpha=0.25$.}
\end{center}
\end{figure}

For $0\leq\alpha<0.23$, the phase diagram seems the ones shown in Figures \ref{alph0} and \ref{alph01}. By contrast, for $\alpha\geq0.23$, 
the phase diagram is modified, namely
only the CAF phase is stabilized along the $\eta$-axis. A phase diagram in this range of frustration parameter is depicted in Figure \ref{alph025}.
Therefore the competition between $J_1$ and $J_3$ is responsible to suppress the ferromagnetic order for $\alpha\geq0.23$, 
since $J_3$ frustrates the $J_1$ coupling. 
We remark that this feature is not observed in the corresponding model with both $J_1$ and $J_3$ antiferromagnetic,
by reason of $J_3$ does not frustrate $J_1$, then, in this case, the N\'{e}el order is enhanced in contrast to the collinear one \cite{grif}.

\begin{figure}
\begin{center}
\includegraphics[angle=0,width=8.5cm]{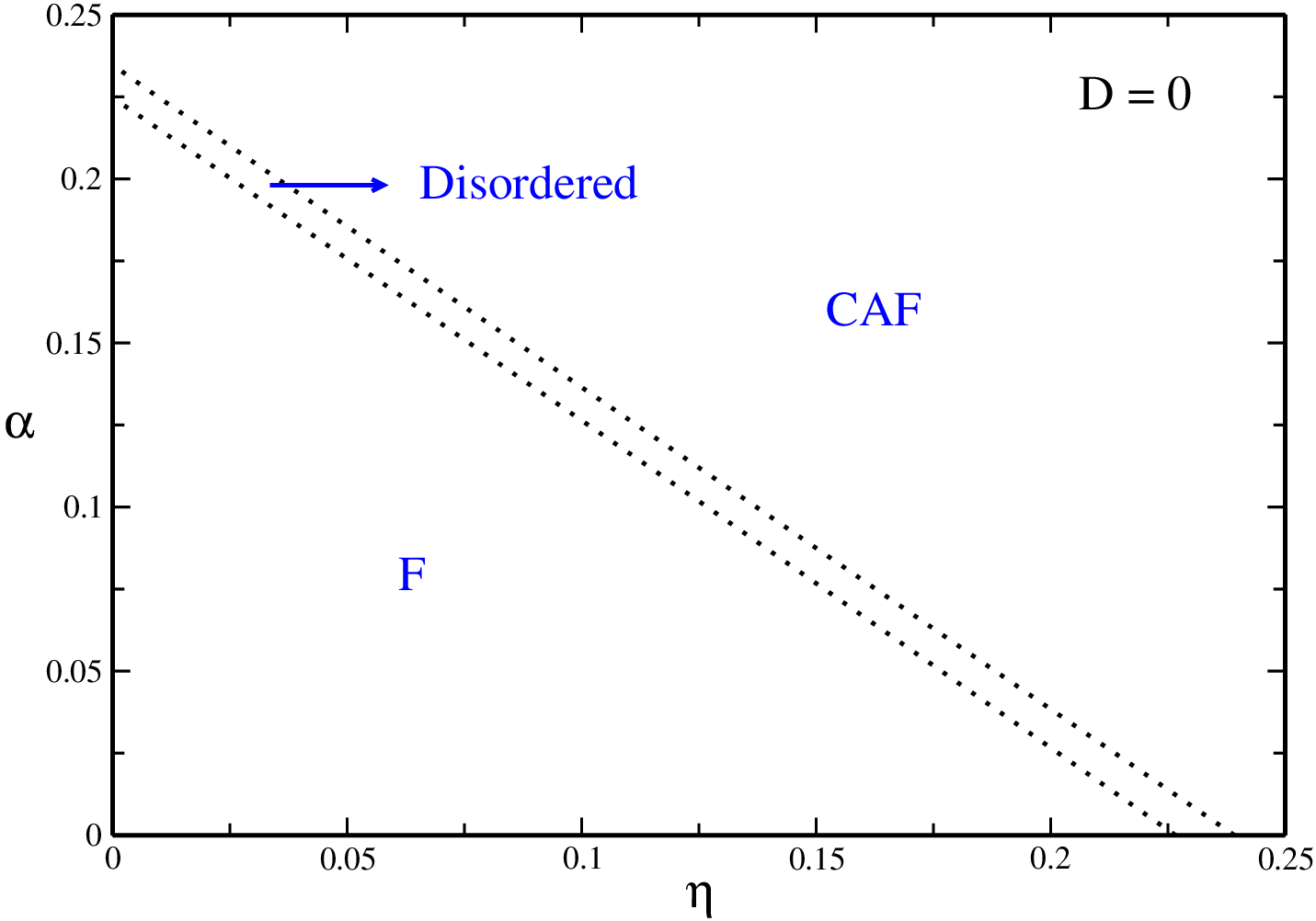} 
\caption{\label{d0} (color online) 
$\alpha$ as a function of $\eta$, at zero temperature, for $D=0$. The dotted lines denote the first-order transitions between the ordered phase (F or CAF) and 
disordered one.}
\end{center}
\end{figure}

For the sake of completeness, the phase diagram for $D=0$ is depicted in Figure \ref{d0}. We show the first-order phase transitions between the non-magnetic phase
(disordered) and the ordered phase (F or CAF). As one can see, the disordered phase is very narrow and its width remains approximately constant as the 
frustration parameters are varied. In addition, we observe that both frustration parameters, $\alpha$ and $\eta$, have the same effect on the phase diagram:
upon increasing $\alpha$ or $\eta$ the phase transitions always occur from the ferromagnetic phase to the disordered phase, and then to the collinear phase.
\vspace{1cm}
\subsection{Phase Diagram at Finite Temperature}
Up to now, we have set $T=0$. In this section, we turn our attention to the effect of temperature on the phase diagram of the present model. Thus,
in contrast to the quantum phase transitions, the phase transitions analyzed here is driven by thermal fluctuations.

\begin{figure}[p]
\begin{center}
\includegraphics[angle=0,width=8.5cm]{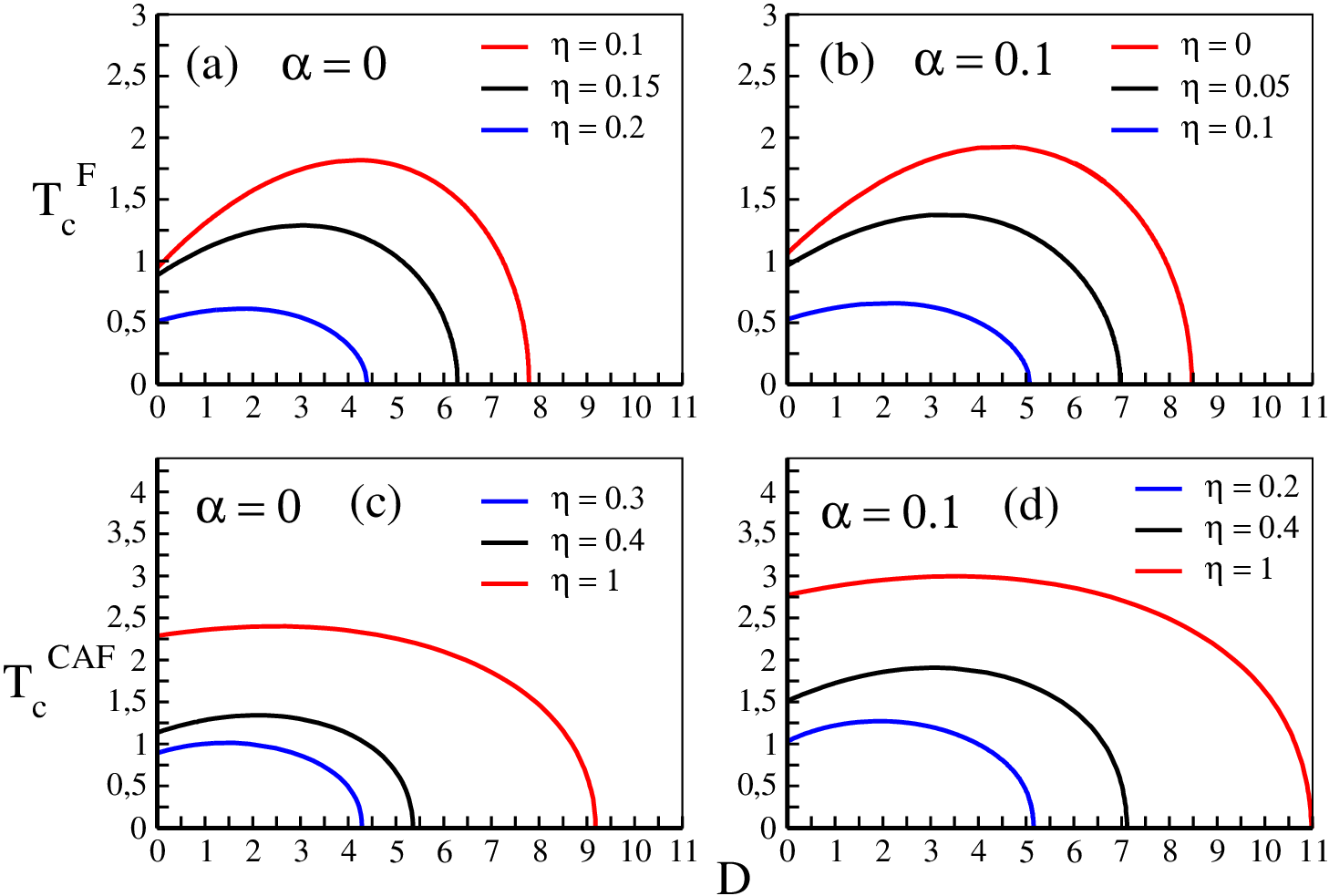} 
\caption{\label{tcXD} (color online) 
Critical temperature, $T^{F(CAF)}_{c}$, as a function of the crystal field, $D$, for several values of frustration parameters, $\alpha$ and $\eta$.}
\end{center}
\end{figure}

The behavior of the critical temperature, ${T_c}^{F(CAF)}$, as a function of the crystal field, for some values of $\eta$, is shown in Figure \ref{tcXD}.
 Phase transitions between the 
F phase and the disordered phase is depicted in Figures \ref{tcXD}(a)-(b) while the transitions between CAF phase and the disordered one is 
depicted in Figures \ref{tcXD}(c)-(d). A common feature is shared by all lines in the Figure \ref{tcXD}: 
the critical temperature passes through a maximum upon increasing
crystal field, and then decreases toward the quantum critical point. 
It should be mentioned that this is not a consequence of frustration, since unfrustrated Heisenberg model also exhibits it
\cite{wang-wang,diego-PRE}. We believe that the quantum nature of the model, which is introduced by the non-commutativity of quantum mechanics spin operators,
is responsible for this effect because in the classical Heisenberg model such characteristic is not observed: the critical temperature increases as
the crystal field increases, and asymptotically approaches a constant value \cite{diego-class}. Consequently, in contrast to its quantum counterpart, 
the spins order down to absolute zero, even if the easy-plane anisotropy is very strong.

In Figures \ref{tcXD}(a)-(b), one notices that ${T_c}^F$ decreases with increasing $\eta$, since the frustration weakens the 
ferromagnetic order, then less thermal fluctuations are required to drive the phase transition. By contrast, in Figures \ref{tcXD}(c)-(d), one can see that
${T_c}^{CAF}$ increases as $\eta$ increases because the collinear order is strengthened by the presence of $J_2$ coupling.
We also conclude, by setting $\eta$, that ${T_c}^F$ is greater for $\alpha=0$ (red line in Fig. \ref{tcXD}(a)) than for $\alpha=0.1$ (blue line in 
Fig. \ref{tcXD}(b)). On the other hand, by doing the same, we observe that ${T_c}^{CAF}$ is greater for $\alpha=0.1$ (red and black lines in Fig. \ref{tcXD}(d)) 
than for $\alpha=0$ (red and black lines in Fig. \ref{tcXD}(c)).
 
\begin{figure}[p]
\begin{center}
\includegraphics[angle=0,width=8.5cm]{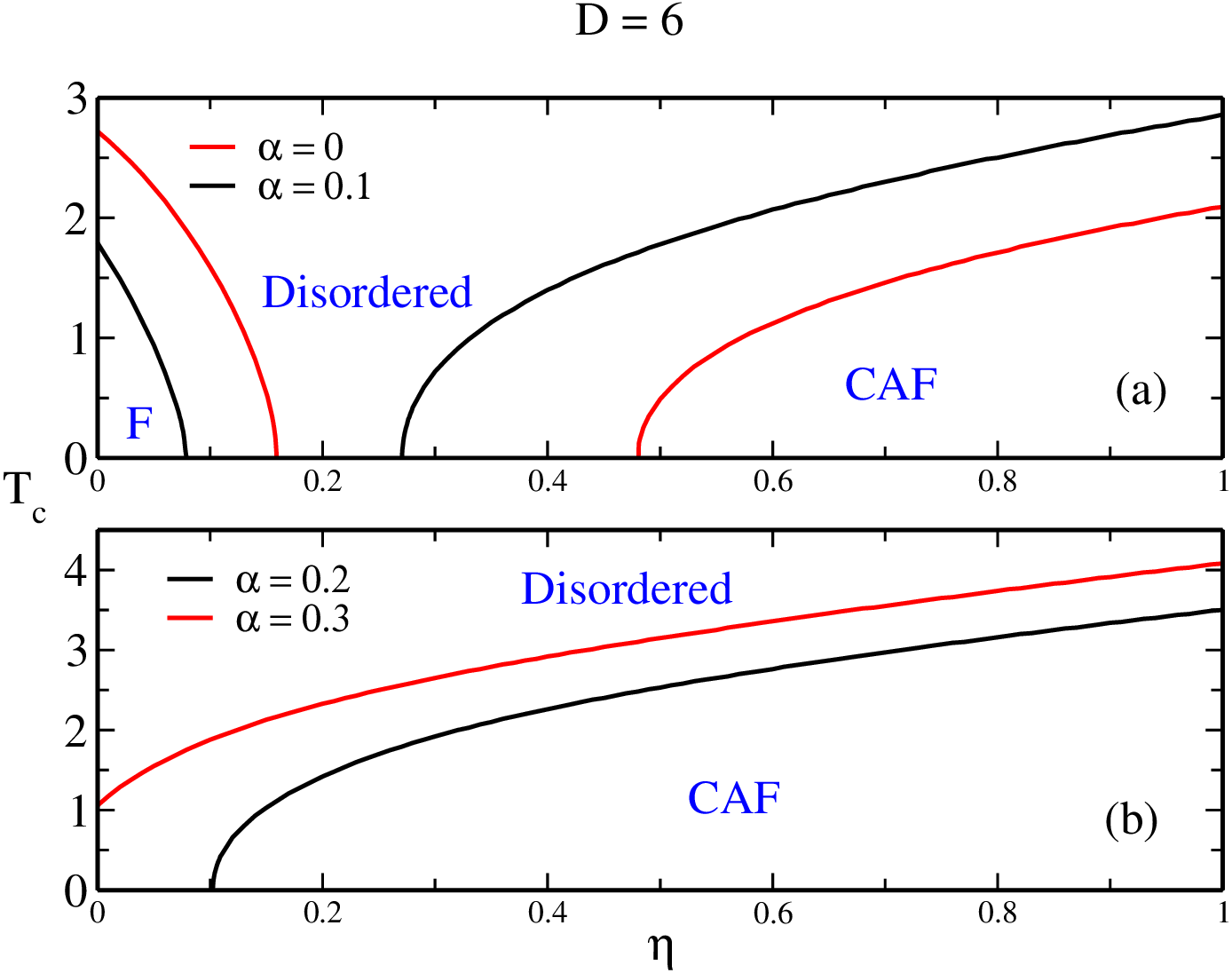} 
\caption{\label{TcXeta} (color online) 
Critical temperature, $T_{c}$, as a function of the frustration parameter, $\eta$, for some values of $\alpha$.
}
\end{center}
\end{figure}

It seems worthwhile to analyze the behavior of the critical temperature as a function of $\eta$ for some values of $\alpha$ with $D\neq0$. The results
that are shown in Figure \ref{TcXeta}(a) for $D=6$ indicate that the magnetically disordered region between F and CAF becomes smaller with increasing $\alpha$. 
For $\alpha\geq0.2$, as depicted in Figure \ref{TcXeta}(b), the F phase is completely destroyed and the CAF region becomes larger.

\section{Concluding Remarks}
\label{con}
%
%
We have studied the  spin-$1$ $J_1-J_2-J_3$ ferromagnetic Heisenberg model with an easy-plane crystal field on the cubic lattice
by using the bond-operator formalism. Phase diagrams have been examined at finite and at zero temperatures. 
The zero-temperature phase diagrams exhibit a narrow magnetically disordered phase between the ferromagnetic and collinear antiferromagnetic phases
for $D=0$ and $0\leq\alpha<0.23$. We suggest that such a disordered phase is a quantum spin liquid candidate. Second- and first-order phase 
transitions have been located according to the energy gap behavior.  
The effect of the crystal field, frustration and temperature on the stability of the ordered phases 
has also been analyzed in some cases. 

One should mention that applications of numerical procedures to the present model would be very welcome in order to be compared with our
analytical results, mainly to confirm the first-order
transitions between the ordered and the disordered phases. Although, in general, numerical methods are not suitable to treat quantum 
systems in three dimensions,
the variational cluster approach has been recently extended to study such systems \cite{vojta}, and then it can be used to verify the nature of these transitions.  

As a final remark, the possibility of defining a nematic order induced by the crystal field ($D\neq0$) at absolute zero in the present model will be subject of
future investigation, since a nematic order has been found in its two-dimensional counterpart \cite{pires}. Furthermore, the extension of the present calculations
to three-dimensional lattices with more complex geometries, such as hyperhoneycomb and stripyhoneycomb \cite{PRL99,tak}, deserves serious consideration.
\section{Acknowledgements}
The authors thank FAPEMIG (Funda\c{c}\~{a}o de Amparo \`{a} Pesquisa do Estado de Minas Gerais) and CNPq (Conselho Nacional de Desenvolvimento Cient\'ifico 
e Tecnol\'ogico) for the financial support. 
\section{References}

\end{document}